\documentclass[twocolumn,pra,showpacs,superscriptaddress,amssymb,amsmath,amsmath,floatfix]{revtex4-1}
\usepackage{graphicx}
\usepackage{epstopdf}
\usepackage{bm}
\usepackage[colorlinks=true,linkcolor=blue,urlcolor=blue,citecolor=blue]{hyperref}
\usepackage{comment}
\usepackage[normalem]{ulem}
\usepackage{float}
\usepackage{cancel}
\usepackage{times}
\usepackage{amsmath, amsthm, amssymb, amsfonts}
\usepackage{xcolor}

\newcommand{\affFUW}{Faculty of Physics, University of Warsaw, Pasteura 5, 02-093 Warsaw, Poland}

\begin{document}
\title{Modified mean field ansatz for charged polarons in a Bose-Einstein condensate}
\author{Ubaldo Cavazos Olivas } 
\email{ubaldo.olivas@fuw.edu.pl}
\affiliation{\affFUW}
\author{Luis A. Pe\~na Ardila}
\email{luis.penaardila@unicam.it}
\affiliation{School of Science and Technology, Physics Division, University of Camerino, Via Madonna delle Carceri, 9B-62032(MC),Italy}
\author{Krzysztof Jachymski}
\email{krzysztof.jachymski@fuw.edu.pl}
\affiliation{\affFUW}
\date{\today}

\begin{abstract}
	
Ionic Bose polarons are quantum entities emerging from the interaction between an ion and a Bose-Einstein condensate (BEC), featuring long-ranged interactions that can compete with the gas healing length, resulting in strong interparticle correlations and enhancement of the gas density around the ion. Full numerical treatment of such systems is computationally very expensive and does not easily allow to study the system dynamics. For this purpose, we study a mean-field based description of such systems in the co-moving frame. We find that it captures a sizeable change in the gas density and qualitatively reproduces the available results based on Monte Carlo simulations. We consider a couple of scenarios which consist of a single and two pinned ions, where it is possible to extract their effective interaction induced by the bath. This approach seems to be promising for studying transport and nonequilibrium dynamics of charged (bi)polarons in condensed media.
\end{abstract}
\maketitle

{\it Introduction ---}. The quantum impurity problem lies at the heart of condensed matter theory. Quasiparticles arising from the impurity-medium interactions, called polarons,  can be considered as a building block for many-body physics~\cite{Alexandrov1992,Franchini2021}, and have been studied since the early days of quantum mechanics~\cite{Landau1948,Lee1953,Frohlich1954,Gross1962}. Nowadays, controllable quantum impurities also offer a wide scope of applications, especially in the context of precision measurements~\cite{Mehboudi2019}.

Ultracold matter is a unique platform allowing for realization of idealized systems with many remarkable experimental results, especially in the context of quantum simulations~\cite{Bloch2012,Monroe2021}. Polarons have been observed and intensely studied in both degenerate Bose and Fermi gases, in the regime of weak as well as strong interactions~\cite{Massignan2014,Jorgensen2016,Ming2016,Ardila2019,Yan2020,Levinsen2021,Will-2021,Jager-2022,Ding2022,Ardila2022,Fuji2022}. Even though the system consists of only a single particle and typically weakly interacting gas, calculation of static polaronic properties such as the energy, residue and spectral function is a nontrivial task. Out of equilibrium dynamics such as polaron formation and transport  phenomena~\cite{Astrakharchik2004,Nielsen2019} are particularly appealing in this context, but also much more challenging to study. Another interesting aspect is the consideration of long-ranged potentials for which the gas can become strongly perturbed and the quasiparticle picture may break down, replaced by formation of a many-body bound state. In particular, ion-atom systems are a prototypical example in which strong and long-ranged interactions among their constituents may occur~\cite{Tomza-2019,Feldker2020,Mohammadi2021,Weckesser2021,Oghittu2021,Dieterle2021,Hirzler2022}. 
Large size of the corresponding potential well can in this case enables bound state occupation by many bosons, leading to a cluster--like many--body bound state~\cite{Casteels2011,Astrakharchik2021,Christensen2021,Christensen-2022,Astrakharchik-2022}. Ionic polarons can thus have vastly different properties from their neutral counterparts and theoretical tools should be applied with care.
Specifically, Bogoliubov theory, although it holds in the weak coupling regime, is typically based on the assumption of a homogeneous condensate and cannot properly capture the strongly inhomogeneous atomic density profile. Instead, one should take a step back and recalculate the bosonic vacuum state over which the Bogoliubov expansion is performed. The applicability of this approach has been shown to work as long as the local gas parameter in the vicinity of the impurity remains small~\cite{Volosniev2017,Yegovtsev2023}. For mobile impurities, a modified mean-field ansatz can be applied in a co-moving frame, which has been shown to yield consistent results. In particular, increased gas density provides a self-stabilization mechanism of the gas with finite-range interactions~\cite{Schmidt-Enss2022}, preventing it from collapse. This method has also revealed universal properties of neutral polarons, as the  polaron size and the correlation functions in this approach turn out to depend only on the scattering length and effective range related to the two-body interaction potential.

\begin{figure}
	\centering
\includegraphics[width=0.9\linewidth]{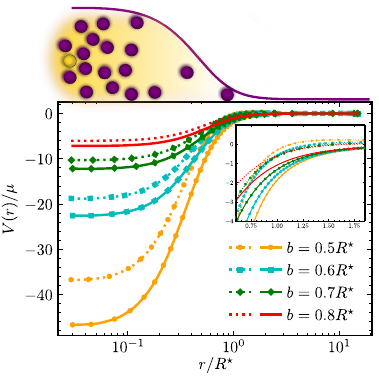}
\caption{\label{fig:effpotc}The bare (solid lines) interaction given by eq.~\eqref{eq:reg_pot} and the effective potential (dotted lines) resulting from eq.~\eqref{eq:eff_pot} for several atom-ion scattering length values with $c = 0.0023R^{\star}$, $n_{0}^{-1/3}=R^{\star}$ and $a_{\mathrm{B}} = 0.0269R^{\star}$. The inset shows the weakly repulsive character of the effective interaction close to $R^\star$. The cartoon illustrates how the buildup of the density of repulsive bosons leads to the dressing of the interaction, with the upper purple (grey) line depicting the repulsive contribution to the effective potential.}
\end{figure}

In this work, we apply this approach to the case of ionic polarons, where the interactions have long-ranged character. Based on earlier numerical results~\cite{Astrakharchik2021}, one can expect that in this case the atoms should be drawn into the potential well, resulting in density increase at length scales comparable to the range of the potential as illustrated in Fig.~\ref{fig:effpotc}. This means that the gas will probe the details of the potential surface and general features such as the screening effect described for short-range potentials can display nonuniversal features.\\


{\it Mean field ansatz ---}
 We start by recalling the derivation of the energy functional that captures the physical properties of our system, following closely~\cite{Gross1962}, and later~\cite{Schmidt-Enss2022}. We are interested in the characterization of a mobile charged impurity surrounded by a weakly interacting Bose-Einstein condensate. The Hamiltonian reads
\begin{equation}
\begin{split}
    H &= \frac{\hat{\boldsymbol{p}}^{2}}{2m_{\mathrm{I}}} + \sum_{i}V_{\mathrm{IB}}(\hat{\boldsymbol{x}}_{\mathrm{B},i}-\hat{\boldsymbol{x}})\\
    &\;\;\;\;\;\; + \sum_{i} \frac{\hat{\boldsymbol{p}}_{\mathrm{B},i}^{2}}{2m_{\mathrm{B}}} + \frac{1}{2}\sum_{i\neq j}V_{\mathrm{BB}}(\hat{\boldsymbol{x}}_{\mathrm{B},i}-\hat{\boldsymbol{x}}_{\mathrm{B},j}),
\end{split}
\end{equation}
where the impurity is described by its mass $m_{\mathrm{I}}$, position $\hat{\boldsymbol{x}}$ and momentum $\hat{\boldsymbol{p}}$, while the mass of identical bosonic atoms is denoted in terms of the two-boson scattering length $a_{\mathrm{BB}}$ as  $m_\mathrm{B}$ with  $\hat{\boldsymbol{x}}_{\mathrm{B},i}$ and $\hat{\boldsymbol{p}}_{\mathrm{B},i}$ being the bosonic position and momentum operators. 
A commonly used simplification may be achieved through a Lee-Low-Pines transformation~\cite{Lee1953} which applies the total momentum conservation to remove the impurity degrees of freedom, and can be thought of as working in a frame co-moving with the impurity. Consequently, using the generator
$S = \exp \Big(i\hat{\boldsymbol{x}}_{\mathrm{}}\,\cdot\sum_{i}\hat{\boldsymbol{p}}_{\mathrm{B},i}\Big)$, one obtains the new Hamiltonian, $H_{\mathrm{LLP}} =  SHS^{-1}$, written as
\begin{equation}\label{LLP-Hamiltonian}
\begin{split}
H_{\mathrm{LLP}} &=  \frac{1}{2m_{\mathrm{I}}}\left(\boldsymbol{p}_{0} - \sum_{i} \hat{\boldsymbol{p}}_{\mathrm{B},i}\right)^{2} + \sum_{i}V_{\mathrm{IB}}(\hat{\boldsymbol{x}}_{\mathrm{B},i})\\
&\;\;\;\;\;\; + \sum_{i} \frac{\hat{\boldsymbol{p}}_{\mathrm{B},i}^{2}}{2m_{\mathrm{B}}} +  \frac{1}{2}\sum_{i\neq j}V_{\mathrm{BB}}(\hat{\boldsymbol{x}}_{\mathrm{B},i}-\hat{\boldsymbol{x}}_{\mathrm{B},j}).
\end{split}
\end{equation}
Here $\boldsymbol{p}_{0} $ is the total momentum of the system. Note now that the impurity coordinates have been removed and the Hamiltonian describes a Bose gas in an external potential $V_{\mathrm{IB}}$ with an additional term coupling the momenta, which has a complicated structure and only disappears for a static impurity. In this work we will restrict the analysis to the case of vanishing $\boldsymbol{p}_{0}$, for which the system has spherical symmetry. Hereafter, we also consider contact interaction among the bosons described by a pseudopotential $V_{\mathrm{BB}}({\boldsymbol{x}}) =g\,\delta^{(3)} (r)$ and its strength using atomic units (i.e. $\hbar=1$) is denoted by $g = 4\pi a_\mathrm{BB}/m_{\mathrm{B}}$. We assume that the ground state is a product of identical symmetric single-particle wave functions
\begin{equation}
\Psi(\boldsymbol{x}_{1}, \boldsymbol{x}_{2}, ..., \boldsymbol{x}_{N} ) = \prod_{j=1}^{N}\phi(\boldsymbol{x}_{j}),
\end{equation}
where $\phi(\boldsymbol{x})$ satisfies the normalization condition $\int d\boldsymbol{x}\big|\phi (\boldsymbol{x})\big|^{2} = N$.
With this product wave function one can calculate the energy of the system, taking into account the normalization constraint,  which leads to a modified mean field (MMF) functional
\begin{equation}
\begin{split}\label{energ_func}
E_{\,\mathrm{1}}[\phi] &=  \int d\boldsymbol{x}\Bigg[\frac{1}{2m_{\mathrm{red,1}}}\big|\boldsymbol{\nabla}\phi (\boldsymbol{x})\big|^{2}+V_{\mathrm{IB}}({\boldsymbol{x}})\big|\phi (\boldsymbol{x})\big|^{2}\\
 &+  \frac{g}{2}\big|\phi (\boldsymbol{x})\big|^{4} -\mu \big|\phi (\boldsymbol{x})\big|^{2} \Bigg]\, .
\end{split}
\end{equation}
This energy functional is described by the chemical potential $\mu = gn_{0}$ with an unperturbed condensate density $n_{0}$, the reduced mass $m_{\mathrm{red,1}}^{-1} = m_{\mathrm{B}}^{-1} + m_{\mathrm{I}}^{-1}$, and the distance from the impurity $r = |\boldsymbol{x}|$. One can now introduce the healing length $\xi = 1/\sqrt{8\pi a_{\mathrm{BB}}n_{0}}$ or, recalling the definition of the coupling constant $g$, $\xi = 1/\sqrt{2m_{\mathrm{B}} \mu}$. Due to the form of the kinetic energy term in~\eqref{energ_func}, it is also useful to define $\xi_{1} = 1/\sqrt{2m_{\mathrm{red,1}}\mu}$. We aim to compute $\phi(\boldsymbol{x})$ which minimizes the energy functional~\eqref{energ_func}, 
fulfilling the boundary conditions $|\phi(\boldsymbol{x}\rightarrow 0)| < \infty$ and $|\phi(|\boldsymbol{x}|\rightarrow \infty)| = \sqrt{n_{0}}$.
The exercise may be simplified if the potential $V_{\mathrm{IB}}(\boldsymbol{x})$ has radial symmetry. As a result, the ground state of the wave function is real and spherically symmetric. Likewise, it is natural to propose an auxiliary radial function $u(r)=r\phi (r)/\sqrt{n_{0}}$ whose corresponding Dirichlet boundary conditions are $u(0) = 0$ and $u(r\rightarrow \infty ) = r$, and a dimensionless energy functional \eqref{energ_func} may be rewritten as \cite{Schmidt-Enss2022}
\begin{equation*}
\begin{split}
\frac{E_{1}[u]}{\mu} &= 4\pi n_0\int_{0}^{\infty} dr\left\{\xi_{1}^{2}\left[\left(\frac{du}{dr}\right)^{2} - 1\right]\right.\\
&\;\;\;\;\;\;\;\;\;\;\;\;\;\;\;\;\;\;\;\;\;\;\;\;\;\; \left. +\frac{1}{\mu}V_{\mathrm{IB}}(r){u(r)}^{2} +  \frac{(u^2-r^2)^2}{2r^2} \right\}.
\end{split}
\end{equation*}
Note that this problem is equivalent to finding a function which satisfies the following differential equation:
\begin{equation}\label{funct-deriv-1d}
\Bigg[ -\xi_{1}^{2} \frac{d^{2}}{dr^{2}}+ \frac{1}{\mu}V_{\mathrm{IB}}(r) +\frac{(u^{2}-r^2)}{r^2}\Bigg]u = 0\, .
\end{equation}
Having the condensate wave function, one can define an effective interaction modified by the gas density as
\begin{equation}
	\label{eq:eff_pot}
	V_{\mathrm{P}}(r) = V_{\mathrm{IB}}(r) + g\left|\phi(r)\right|^{2} - \mu\, .
\end{equation}
This effective potential is more repulsive than the bare one due to the mean-field correction, which can prevent the Bose gas from collapsing in the vicinity of the impurity~\cite{Schmidt-Enss2022}.\\

{\it Bipolaron formation ---}
Due to the local deformation of the gas density, impurities are prompted to attract each other which can lead to formation of bipolarons. 
We therefore neglect the direct interaction between the two particles to focus on the bath-induced properties. This can be justified by assuming that the ions are placed in an external trap, which fixes the distance between them and balances out the Coulomb interaction~\cite{James1998}. Their motion is then conveniently represented by the center of mass $\hat{r}$ and relative position among the impurities $\hat{R}$ and their corresponding momenta $\hat{p}$ and $\hat{P}$, respectively.  Once more, we start in the laboratory frame and perform a Lee-Low-Pines transformation, removing the center of mass coordinate. Consequently, the Hamiltonian is \cite{Will-2021, Jager-2022},
\begin{equation}
\begin{split}
H_{\mathrm{LLP}} &=  \frac{1}{4m_{\mathrm{I}}}\left(\boldsymbol{p}_{0} - \sum_{i} \hat{\boldsymbol{p}}_{\mathrm{B},i}\right)^{2} + \frac{\hat{P}^2}{m_\mathrm{I}} + \sum_{i}\sum_{\alpha=\pm}V_{\mathrm{IB}}(\hat{\boldsymbol{x}}_{\mathrm{B},i\alpha})\\
&\;\;\;\;\;\;\;\;\;\;\;\;\;\; + \sum_{i} \frac{\hat{\boldsymbol{p}}_{\mathrm{B},i}^{2}}{2m_{\mathrm{B}}} +  \frac{1}{2}\sum_{i\neq j}V_{\mathrm{BB}}(\hat{\boldsymbol{x}}_{\mathrm{B},i}-\hat{\boldsymbol{x}}_{\mathrm{B},j}),
\end{split}
\end{equation}
where  $\boldsymbol{x}_{\mathrm{B},i} = \boldsymbol{r}_{i}$ and the relative distance between the impurities and bosons $\boldsymbol{x}_{\mathrm{B},i\,\pm} = \boldsymbol{r}_{i} \pm \boldsymbol{R}/2$. Analogously to the single-impurity case, a product state ansatz is adopted and we consider a cylindrically symmetric potential. Using symmetry arguments and fixing the relative distance between the ions, treating it as a parameter in the spirit of the Born-Oppenheimer approximation and taking $\boldsymbol{p}_{0} = 0$, the terms involving the kinetic energy are negligible. As a result, one finds the following energy functional:
\begin{equation}\label{bipolaron_energy}
\begin{split}
\frac{E_{2}(\phi, \boldsymbol{R})}{\mu} &=\int d^{3} r\left\{\xi_{2}^{2}\big|\boldsymbol{\nabla}\phi\big|^{2} +  \frac{n_0}{2}(\big|\phi \big|^{2}/n_{0}- 1)^2\right.\\
&\;\;\;\; +\, \left.\left[V_{\mathrm{IB}}(\boldsymbol{r} + \boldsymbol{R}/2) + V_{\mathrm{IB}}(\boldsymbol{r} - \boldsymbol{R}/2)\right]\big|\phi \big|^{2}/\mu \right\}.
\end{split}
\end{equation}
Notice that here we have used $\xi_{2} = 1/\sqrt{2m_{\mathrm{red,2}}\mu}$ which corresponds to a reduced mass $m_{\mathrm{red,2}}^{-1} = m_{\mathrm{B}}^{-1} + (2m_{\mathrm{I}})^{-1}$. The minimizer of Eq.~\eqref{bipolaron_energy} satisfies a boundary value problem, described by the partial differential equation
\begin{equation}\label{pde_bipolaron}
\begin{split}
\left[-\xi_{2}^{2}\boldsymbol{\nabla}^{2} + \left(V_{\mathrm{IB}}(\boldsymbol{r} + \boldsymbol{R}/2) + V_{\mathrm{IB}}(\boldsymbol{r} -  \boldsymbol{R}/2)\right)/\mu\right.\\
\left. +\,  (\big|\phi (r,z)\big|^{2}/n_0-1)\right]\phi (r,z) = 0,
 \end{split}
\end{equation}
with Dirichlet boundary conditions $
\phi (r\rightarrow\infty, z) = \phi (r, z\rightarrow -\infty ) =  \phi (r, z\rightarrow\infty) = \sqrt{n_{0}}$.
One can then extract the induced interaction between the impurities mediated by the condensate by subtracting the contribution to the system energy from single polarons
$
V_{\mathrm{BP}}(R) = E_{2}(R) -  2E_{1} +E_0$, where $E_{0}$ is the energy of the unperturbed Bose gas that we set to zero here.\\

{\it Ion-atom interactions ---}
Let us now briefly discuss the properties of the ion-boson interaction mediated by the potential $V_{\mathrm{IB}}(r)$. The leading contribution at large distances comes from the charge-induced dipole interaction~\cite{Tomza-2019} $
V_{\mathrm{IB}}(r\longrightarrow\infty) = -\frac{C_{4}}{r^{4}}$,
where the induction coefficient $C_{4} =\alpha q^{2}/2$, $\alpha$ being the static electric dipole polarizability of the atom and $q$ the ion charge. It leads to a characteristic length $R^{\star} = \sqrt{2m_{\mathrm{red,1}}C_4}$ and energy~$E^{\star} =1/[2m_{\mathrm{red,1}}(R^{\star})^2]$. In order to stabilize its short-range behavior, we use the regularized potential \cite{Krych-2015}:
\begin{equation}
	\label{eq:reg_pot}
V_{\mathrm{IB}}(r) = -\frac{C_{4}}{(r^{2} + b^{2})^{2}}\frac{r^{2} - c^{2}}{r^{2} + c^{2}}.
\end{equation}


\begin{figure*}
	\centering
    \includegraphics[width=0.4\linewidth]{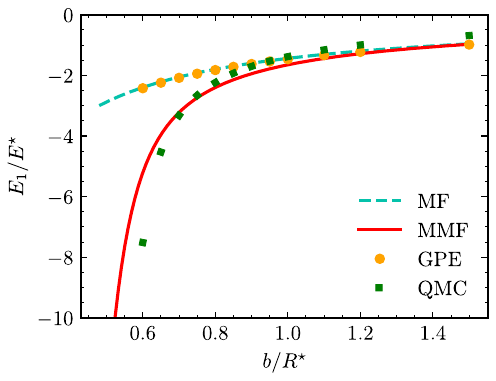}
    \includegraphics[width=0.4\linewidth]{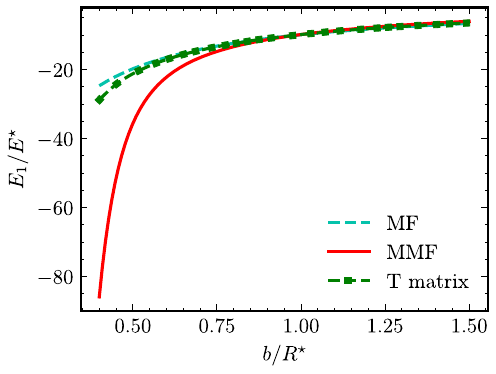}
	\caption{Left: energy of a single polaron in units of $E^\star = 1/[m_{\mathrm{B}}(R^{\star})^{2}]$ as a function of the regularization parameter $b$, with $n_0=0.1458\,(R^\star)^{-3}$. Results of the modified mean field theory in the co-moving frame (straight red (grey) line) are compared with QMC calculation (green (gray) dots) as well as standard mean-field theory where $E_1=-\pi^2 n/b$ (light blue (light grey) dashed) and the GPE calculation (yellow (ligt gray) dots). Right: results for higher density $n_0=(R^\star)^{-3}$ with the T-matrix results of~\cite{Christensen2021} (green (gray) dashed with dots) and the mean-field prediction (light blue (light grey) dashed) for comparison.}
	\label{fig:single_polen}
\end{figure*}

The parameter $b$ controls the potential depth, while $c$ provides a bound for the distance at which the interaction becomes repulsive. This parametrization allows for controlling the scattering length as well as the number of bound states mediated by the potential. At $b\gtrsim 0.58$ the potential is so shallow that it does not have a bound state, and as a result the scattering length is small and negative.\\


{\it Results ---}
Throughout this section, we present the static properties of the polaron and bipolaron described above. In order to obtain these results, 
we have numerically solved the differential equations for the condensate wave function~\eqref{funct-deriv-1d} and \eqref{pde_bipolaron} using ~\texttt{FEniCS} software~\cite{Logg-2012}.  For comparison, we have also calculated the polaronic properties solving the full 3D Gross-Pitaevskii equation with a mobile as well as pinned impurity. We choose to work with equal boson and impurity masses, having in mind e.g. the experimental realization using $^{87}$Rb/Rb$^+$ system~\cite{Dieterle2021}. For infinitely large $m_\mathrm{I}$ the impurity becomes pinned and the reduced mass is just the boson mass, making the Lee-Low-Pines transformation obsolete.

The single impurity case is essentially one-dimensional. Under such circumstances we use the simplest arrangement for a 1D mesh, i.e. a linear rod, with 100 grid points. The finite element scheme implemented in \texttt{FEniCS}  that we utilize consists of a uniformly partitioned mesh, followed by space discretization using second order Lagrange polynomials.
In the ultradilute limit $\xi_{1}\gg R^{\star},n_0^{-1/3}$ we expect no significant effect of the $r^{-4}$ interaction. Therefore, we work at the values of condensate density $n_{0}$ comparable to $\xi_{1}^{-3}$, and assume the competing length scenario in which also $\xi_{1}\approx R^{\star}$. In order to vary between different regimes of the interaction, we set the parameter~$c = 0.0023R^{\star}$, crucial for a repulsive potential at $r\lesssim a_{0}~$\cite{Astrakharchik2021}. Then by changing the $b$ parameter, we can achieve a shallow or deep potential with or without a bound state with tunable scattering length. 
 
The minimization of the MMF functional~\eqref{energ_func} yields the ground state energy of the system which we show in Fig.~\ref{fig:single_polen} and compare to the ladder approximation result of~\cite{Christensen2021} using exactly the same parameters. The two approaches agree well in the limit of very weak interactions $b\gtrsim 1$ as expected. For deeper potentials, the MMF curve bends towards lower energies due to the buildup of atomic density around the ion which increases the integrated interaction energy. The standard mean field formula $E_1=4\pi na/m$~\cite{Scazza2022} with $a$ being the ion-atom scattering length lies very closely to the ladder approximation result as it also assumes a homogeneous gas. We also include a comparison with a QMC calculation for 192 bosons at a lower density $n_0=0.1458\,(R^\star)^{-3}$. The MMF and QMC agree very well in the whole range of $b$ considered, while standard mean field theory leads to a higher estimation of energy. Also the solution of standard GPE equation for two coupled fields representing the ion and the atoms, shown as yellow dots, can only reproduce the QMC and MMF results for large values of $b$ where essentially all approaches agree with each other. 

\begin{figure}
	\centering
	\includegraphics[width=0.7\linewidth]{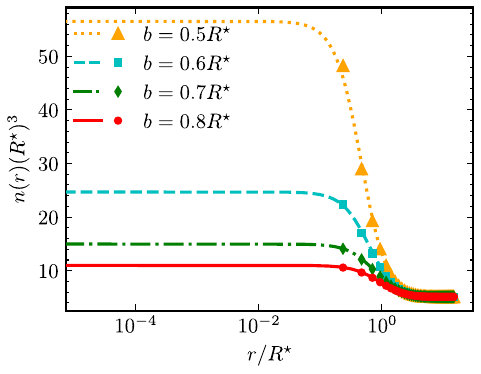}
        \includegraphics[width=0.7\linewidth]{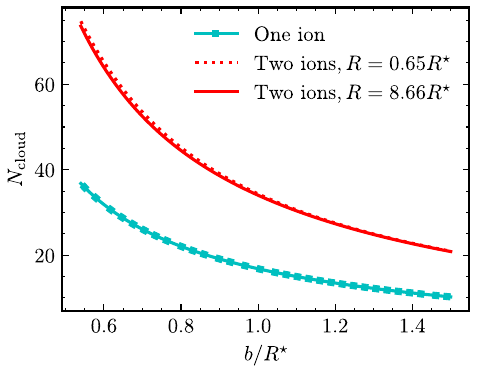}
	\caption{Upper: the gas density as a function of the distance from the impurity $r$ for several values of $b$. The dots indicate the results of solving the functional in full three-dimensional space. Lower: the number of bosons in the polaronic cloud as a function of the potential regularization parameter $b$ for one ion (light blue (grey) with dots) and two pinned ions at two different distances $R$ (straight and dashed red (grey) lines). The other parameters are the same as in Fig.~\ref{fig:effpotc}.}
	\label{fig:dens-corr-1d}
\end{figure}
The formation of the cloud in the vicinity of the ion modifies the effective potential~\eqref{eq:eff_pot} as a result of the mean field boson repulsion, as illustrated in Fig.~\ref{fig:effpotc}. In addition to becoming more shallow, we note that the potential loses its power-law decay and even becomes weakly repulsive at intermediate distances, but retains a similar range. Furthermore, the gas density profile is given by $u(r)$, which can also be interpreted as the impurity-boson correlation function $g^{(2)}_{\mathrm{IB}}(r) = \left|\phi(r)\right|^{2}/n_{0}$. It exhibits a smooth shape which peaks close to the impurity and asymptotically flattens to an unperturbed BEC for distances far away from the ion. In Fig.~\ref{fig:dens-corr-1d}, the behavior of the gas density is depicted for various $b$ values in the weak interaction regime. The peak density value can be an order of magnitude larger than the background density. We have checked that for a short-ranged potential with the same scattering length a much lower enhancement of the density around the impurity is obtained. The large range of the interactions can thus lead to new effects even at the mean field level. This feature also manifests itself as clustering of particles in a cloud, which can be quantified as~\cite{Massignan2005}
\begin{equation}
N_{\mathrm{cloud}} =  \int d^{3}r\left[\left|\phi\right|^{2} - n_{0}\right].
\end{equation}
The lower panel of Fig.~\ref{fig:dens-corr-1d} shows the number of bosons trapped by the ion as a function of the regularization parameter $b$. With increasing $b$ the potential becomes more shallow, such that less atoms can fit in the potential well. We observe that this number turns out to be of the order of few tens of atoms, qualitatively similar to the Gaussian potential predictions~\cite{Schmidt-Enss2022}. Furthermore, within the semiclassical approximation~\cite{Massignan2005} one obtains $N_{\mathrm{cloud}}=-2a/a_{\mathrm{BB}}$ which is of the order of a hundred of atoms, again in qualitative agreement. On the other hand, perturbative mean field result for the number of bound particles~\cite{Scazza2022} $\Delta N\approx -a/a_{\mathrm{BB}}-4\sqrt{2}a^2/\pi \xi a_{\mathrm{BB}}$ renders a negative value as $a/\xi$ is not a small parameter here.

\begin{figure*}
	\centering
         \includegraphics[width=0.33\linewidth]{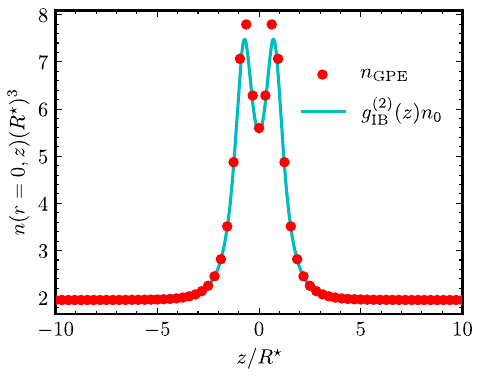}
	\includegraphics[width=0.33\linewidth]{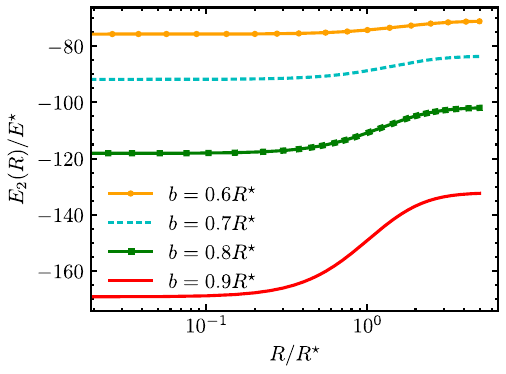}	
	\includegraphics[width=0.33\linewidth]{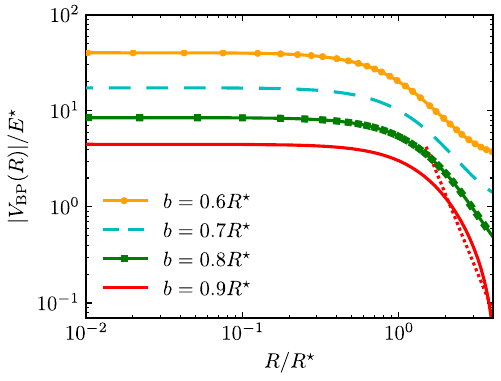}
	\caption{Left: Cut of the gas density profile along $z$ for $r=0$ for two pinned ions. Here
  $\xi_{2}=R^{\star}$, $c = 0.0023R^{\star}$, $b = 0.572R^{\star}$, and $n_{0} = 1.9531 (R^\star)^{-3}$. The red dots indicate the results of a full three-dimensional calculation. Middle: absolute value of the system energy in units of $E^\star$ for two ions as a function of their separation for the same parameters and several $b$ values, right: the resulting induced effective interaction potentials. Dashed line indicates the power law derived within the ladder approximation ~\cite{Ding2022} described by eq.~\eqref{eq:bruun}.}
	\label{fig:bip_merged}
\end{figure*}

In the case of two impurities we set up a 2D rectangular mesh, composed of 100 elements in each dimension, and second-order Lagrange polynomials for space discretization as well. Note that if we assumed $\xi_1=R^\star$, the corresponding value for the healing length is now $\xi_{2} = (\sqrt{3}/2)R^{\star}$ for the same condensate density $n_0$.  The impurity-impurity-boson correlation density profiles, depicted in~Fig.~\ref{fig:bip_merged}, reach a maximum value where the impurities are located. 
We further observe that the number of bound bosons is roughly twice the single-impurity case for both potentials, as presented by the red curve in the lower panel of~Fig.~\ref{fig:dens-corr-1d}. This would be expected if the induced interactions are relatively weak such that the system can be regarded as two almost independent polarons. Indeed, inspecting the density profile shows that the gas density is quite well approximated by a sum of two polaron peaks. We also show in~Fig.~\ref{fig:bip_merged} the energy of the system as well as the effective interaction induced by the medium. We observe that the effective interaction follows a power law at long range and levels off at distances $\lesssim R^\star$, where the buildup of the density starts being significant. The magnitude of the effective potential of the order of $E^\star$ is in qualitative agreement with the numerical results~\cite{Astrakharchik-2022} for the weakly interacting polaron regime. Note that within the ladder approximation, the induced interaction at large distance is given by~\cite{Ding2022}
\begin{equation}
V_{\rm ind}(r)/E^\star=-\frac{\pi}{4}\frac{1}{a_{\mathrm{BB}}b}\frac{b^2+2bc-c^2}{(b+c)^2}\frac{1}{r^4}\, .
\label{eq:bruun}
\end{equation}
This formula displays a weak dependence on the value of the $b$ parameter and agrees with our results best for the largest $b$ values corresponding to the most shallow potential. 

\begin{figure}
	\centering
    \includegraphics[width=0.9\linewidth]{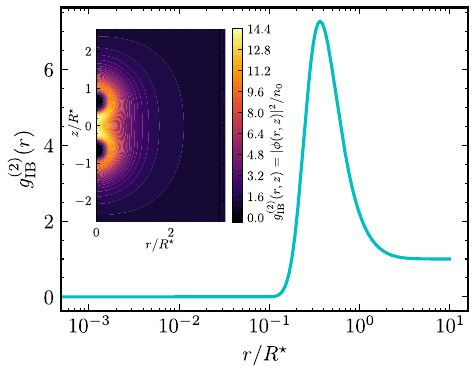}
	\caption{Cut through the atomic density profile for a single impurity for the interaction potential with large short-range repulsive barrier with $\xi_{1}=R^{\star}$,  $b = 0.02 R^{\star}$, $c = 0.225R^{\star}$ and $n_{0} = (0.8R^{\star})^{-3}$. Inset: the density for two impurities separated by $R = 1.3 R^{\star}$ for the same parameters. Results were normalized by the asymptotic atomic density such that $g_{\rm IB}(r)=n(r)/n_0$.}
	\label{fig:dens_comp}
\end{figure}

 
	\label{fig:effpot_res}

The adopted approach relies on a modified mean field ansatz which takes into account the boson repulsion. However, this approximation neglects higher order bosonic correlations and cannot fully reproduce the phenomenon of many-body bound state formation~\cite{Astrakharchik2021}. In order to test the method outside of its expected validity range, we applied it for interaction potential parameters $c=0.225\,R^\star$ and $b=0.02\,R^\star$ for which there appears a strong repulsive barrier at $r\lesssim 0.1\, R^\star$. The resulting density profiles are shown in Fig.~\ref{fig:dens_comp}. As expected, the gas density vanishes close to the ion(s) and the correlation peak is located at a finite distance. However, its maximum value is of the order of five, while the Monte Carlo results give an order of magnitude larger result~\cite{Astrakharchik2021}. This is due to macroscopic occupation of the few-body bound states which is not present in our model as the mean field approach does not resolve individual particles. For the bipolaron case we observe a similar behavior of the peaks in the correlation function being typically much lower than the QMC results~\cite{Astrakharchik-2022}. However, in the weakly interacting regime the two approaches provide qualitatively similar results. Indeed, for the $b=R^\star$ case studied in~\cite{Astrakharchik-2022} an enhancement in density by about a factor of two has been observed, in agreement with the present results.


{\it Conclusions}.
In conclusion, we have studied the ground state properties of charged polarons within the mean field approximation which in the weak coupling limit corresponding to the regularization parameter $b\gtrsim R^\star$ agree qualitatively with more elaborate numerical treatment. For two pinned ions, the method also provides reasonable cloud density profiles and induced interactions. This makes it promising for several future extensions. For instance, switching from static to the dynamical scenario~\cite{Ardila21,Seetharam2021}, as well as including external traps and nonzero total momentum will enable the study of the impurity effective mass and other transport properties, while Monte Carlo-based methods require a lot more intense numerical effort to probe nonequilibrium situations. Note that the considered limit of long range but weak potential cannot describe the creation of many-body bound states, making the system resemble the case of neutral impurity, but with notable quantitative differences in the cloud size and density. It would thus be interesting to enrich the current ansatz to include bosonic correlations explicitly by means of Jastrow wave function~\cite{Drescher2020}, which may extend the validity of the method to larger coupling strength and resonant interactions within the bath.

{\it Acknowledgments}. We acknowledge fruitful discussions with Katarzyna Nurowska, Gregory Astrakharchik, Tilman Enss, Antonio Negretti, Krzysztof My\'sliwy, and Tomasz Wasak. This work was supported by the Polish National Agency for Academic Exchange (NAWA) via the Polish Returns 2019 programme. LAPA acknowledges financial support from PNRR MUR project PE0000023-NQSTI.

\bibliography{refs}

\end{document}